# Alignment, Orientation, and Coulomb Explosion of Difluoroiodobenzene Studied with the Pixel Imaging Mass Spectrometry (PImMS) Camera

Kasra Amini[1], Rebecca Boll[2], Alexandra Lauer[1], Michael Burt[1], Jason W L Lee[1], Lauge Christensen[3], Felix Brauβe[4], Terence Mullins[5], Evgeny Savelyev[2], Utuq Ablikim[6], Nora Berrah[7], Cédric Bomme[2], Stefan Düsterer[2], Benjamin Erk[2], Hauke Höppner[2,8], Per Johnsson[9], Thomas Kierspel[5,10], Faruk Krecinic[4], Jochen Küpper[5,10,11], Maria Müller[12], Erland Müller[2], Harald Redlin[2], Arnaud Rouzée[4], Nora Schirmel[2], Jan Thøgersen[3], Simone Techert[2,13,14], Sven Toleikis[2], Rolf Treusch[2], Sebastian Trippel[5,10], Anatoli Ulmer[12], Joss Wiese[5], Claire Vallance[1], Artem Rudenko[6], Henrik Stapelfeldt[3], Mark Brouard[1], Daniel Rolles[6,2]

[1]*The Chemistry Research Laboratory, Department of Chemistry, University of Oxford, Oxford OX1 3TA, United Kingdom*
[2]*Deutsches Elektronen-Synchrotron DESY, 22607 Hamburg, Germany*
[3]*Department of Chemistry, Aarhus University, 8000 Aarhus C, Denmark*
[4]*Max-Born-Institut für nichtlineare Optik und Kurzzeitspektroskopie, 12489 Berlin, Germany*
[5]*Center for Free-Electron Laser Science (CFEL), Deutsches Elektronen-Synchrotron DESY, 22607 Hamburg, Germany*
[6]*J.R. Macdonald Laboratory, Department of Physics, Kansas State University, Manhattan, KS 66506, USA*
[7]*Department of Physics, University of Connecticut, Storrs, CT 06269, USA*
[8]*Institut für Physik, Carl von Ossietzky Universität, 26111 Oldenburg, Germany*
[9]*Department of Physics, Lund University, 22100 Lund, Sweden*
[10]*Center for Ultrafast Imaging, Universität Hamburg, 22761 Hamburg, Germany*
[11]*Department of Physics, Universität Hamburg, 22761 Hamburg, Germany*
[12]*Institut für Optik und atomare Physik, Technische Universität Berlin, 10623 Berlin, Germany*
[13]*Max Planck Institute for Biophysical Chemistry, 33077 Göttingen, Germany*
[14]*Institute for X-ray Physics, Göttingen University, 33077 Göttingen, Germany*

**Abstract**

Laser-induced adiabatic alignment and mixed-field orientation of 2,6-difluoroiodobenzene ($C_6H_3F_2I$) molecules are probed by Coulomb explosion imaging following either near-infrared strong-field ionization or extreme-ultraviolet multi-photon inner-shell ionization using free-electron laser pulses. The resulting photoelectrons and fragment ions are captured by a double-sided velocity map imaging spectrometer and projected onto two position-sensitive detectors. The ion side of the spectrometer is equipped with the Pixel Imaging Mass Spectrometry (PImMS) camera, a time-stamping pixelated detector that can record the hit positions and arrival times of up to four ions per pixel per acquisition cycle. Thus, the time-of-flight trace and ion momentum distributions for all fragments can be recorded simultaneously. We show that we can obtain a high degree of one- and three-dimensional alignment and mixed-field orientation, and compare the Coulomb explosion process induced at both wavelengths.

# 1 Introduction

Ultrafast lasers provide opportunities to image molecular dynamics taking place on the femtosecond timescale [*Baumert 1991, Zewail 2000, Chergui 2009*]. Table-top Ti:Sapphire laser systems are the most commonly used ultrafast laser systems, producing radiation in the near-infrared (NIR) range. High-intensity femtosecond NIR pulses can rapidly remove several valence electrons from a molecule, producing a multiply charged molecular ion that explodes due to the Coulomb repulsion between its components. The resulting recoil velocities and directions of the product ions depend on the position of the atoms in the molecule before ionization, and consequently can provide structural information about the molecule [*Vager 1989, Stapelfeldt 1995, Posthumus 1996, Hishikawa 1998, Sanderson 1999*]. They can also be used to determine the orientation of molecules in the laboratory frame, for example, to probe the degree of molecular alignment induced by intense laser fields [*Stapelfeldt 2003*], or to probe structural changes of the molecule in time-resolved experiments [*Legare 2005, Hishikawa 2007, Matsuda 2011, Bocharova 2011, Ibrahim 2014, Christensen 2014*].

Absorption of extreme ultraviolet (XUV) and soft X-ray photons can also induce Coulomb explosion when the resulting inner-shell ionization is followed by an Auger process that leads to a multiply charged molecular ion [*Muramatsu 2002, Ueda 2005, Ullrich 2012, Erk 2014, Murphy 2014, Ablikim 2016, Ablikim 2017*]. Free-electron lasers (FELs) produce extremely intense (>$10^{12}$ photons/pulse) and ultrashort (few to few hundred fs) pulses of XUV and X-ray radiation [*Ackermann 2007, Shintake 2008, Emma 2010, Allaria 2012, Ishikawa 2012*], unlocking opportunities to probe ultrafast processes in gas-phase molecules through time-resolved Coulomb explosion imaging experiments [*Johnsson 2009, Jiang 2010, Ullrich 2012, Schnorr 2013, Rouzee 2013, Erk 2014, Schnorr 2014, Fang 2014, Rudenko 2015, Liekhus 2015, Picon 2015, Lehmann 2016, Boll 2016*]. A good understanding of the Coulomb explosion of polyatomic molecules upon absorption of one or several XUV FEL photons in comparison to the Coulomb explosion induced by intense NIR laser fields is therefore essential to further understand time-dependent pump-probe studies, which use XUV FEL photons to probe structural changes in a molecule. Such a comparison can also provide insight into the details of the two different fragmentation processes.

Here, we use Coulomb explosion imaging to study the laser-induced adiabatic alignment and mixed-field orientation of 2,6-difluoroiodobenzene (DFIB, $C_6H_3F_2I$, see inset in Fig. 1) and compare the Coulomb explosion induced by multi-photon inner-shell ionization in the extreme ultraviolet range to that induced by intense femtosecond near-infrared laser pulses within the same experimental setup. The experiment was performed using a doubled-sided velocity map imaging spectrometer [*Strüder 2010, Rolles 2014*] for simultaneous detection of the electron and ion momentum distributions. The application of the Pixel Imaging Mass Spectrometry (PImMS) camera [*Nomerotski 2010, John 2012, Sedgwick 2012, Brouard 2012, Amini 2015*] on the ion side allowed the simultaneous detection of the position and arrival time of up to four charged particles per pixel and per acquisition cycle, therefore providing the momentum distributions of all ions produced in a given laser/FEL shot within a single measurement. Because the FEL has a low repetition rate but high pulse energy, which means that a large number of ions are produced within a single FEL shot, the combination of the PImMS camera with a VMI spectrometer offer experimental capabilities that are hard to achieve otherwise.

We note that DFIB (both the 2,6- and the 3,5-isomers) has been used as a target molecule in several laser-induced molecular alignment and orientation experiments [*Viftrup 2007, Nevo 2009, Ren 2014*]. 2,6-DFIB has a very high polarizability of $\alpha_{zz}$ = 21.3 Å$^3$, $\alpha_{yy}$ = 14.5 Å$^3$, and $\alpha_{xx}$ = 8.5 Å$^3$, where $\alpha_{zz}$, $\alpha_{yy}$, and $\alpha_{xx}$ are the diagonal elements of the polarizability tensor, the z-axis is parallel to the C–I axis, the y-axis is perpendicular to the z-axis but still in the molecular plane, and the x-axis is perpendicular to the molecular plane [*Nevo 2009*]. The fragmentation of 2,6- and 3,5-DFIB after inner-shell ionization has recently been studied with synchrotron radiation [*Ablikim 2017*], as has the dissociation of 3,5- and 2,4-DFIB at various UV-wavelengths [*Murdock 2012*]. In addition to the experiments described here, we have also performed UV-pump XUV-probe experiments, which are reported in a separate publication [*Savelyev 2017*].

## 2   Experimental Setup

The experiment was performed on the focused branch of beamline BL3 at the FLASH free-electron laser [*Feldhaus 2010*] at DESY in Hamburg, Germany. A sketch of the experimental set-up is shown in Figure 1. A detailed description of the set-up is given in [*Savelyev 2017*] and it is therefore only summarized here. A pulsed Even-Lavie valve produced a cold molecular beam of DFIB seeded in neon at a backing pressure of 20 bar. To increase the degree of molecular alignment and orientation achievable, a strong inhomogeneous static electric field provided by an electrostatic deflector [*Filsinger 2009, Küpper 2014, Chang 2015*] was used to spatially disperse the beam and thereby separate the DFIB molecules from the carrier gas and to select molecules in the lowest rotational quantum states [*Holmegaard 2009, Nevo 2009*]. The cold beam was then crossed at 90° by two co-propagating laser beams: the molecules were laser-aligned or mixed-field oriented using a non-resonant linearly or elliptically polarized pulse from a Nd:YAG laser ($\lambda$ = 1064 nm, pulse duration: 12 ns, pulse energy: 1.2 J, estimated spot size: 50 µm (FWHM)). For a linearly polarized pulse, the induced dipole moment and the resulting torque induced by the laser field resulted in one-dimensional (1D) alignment of the most polarizable axis of the molecule (parallel to the C–I axis), whereas three-dimensional (3D) alignment was achieved using elliptically polarized light [*Larsen 2000, Stapelfeldt 2003, Nevo 2009*]. Furthermore, molecular orientation was induced by the combined action of the laser pulse and a weak static electric field [*Holmegaard 2009, Filsinger 2009, Nevo 2009*]. The aligned or oriented molecules were then probed by either an intense 70 fs (FWHM) NIR laser pulse (linearly polarized, $\lambda$ = 800 nm, $I_{probe}$ ~ 2 x 10$^{14}$ W cm$^{-2}$) from the FLASH pump-probe laser [*Redlin 2011*], or by an intense ~100 fs (FWHM) FEL pulse (horizontally linearly polarized, $\lambda$ = 11.6 ± 0.1 nm, corresponding to a photon energy of 107 eV; average pulse energy 32 μJ ± 8 μJ (FWHM), corresponding to an intensity of ~ 3 x $10^{13}$ W $cm^{-2}$ on target, assuming a beamline transmission of 60%). The FEL intensity was determined by the highest pulse energy that was available at the requested pulse duration, while the NIR intensity was set such that a similar ratio of $I^+$ and $I^{2+}$ ion yields was observed in the FEL and NIR data in order to study comparable Coulomb explosion conditions. The lasers and the FEL were operated at a repetition rate of 10 Hz and were electronically synchronized to each other.

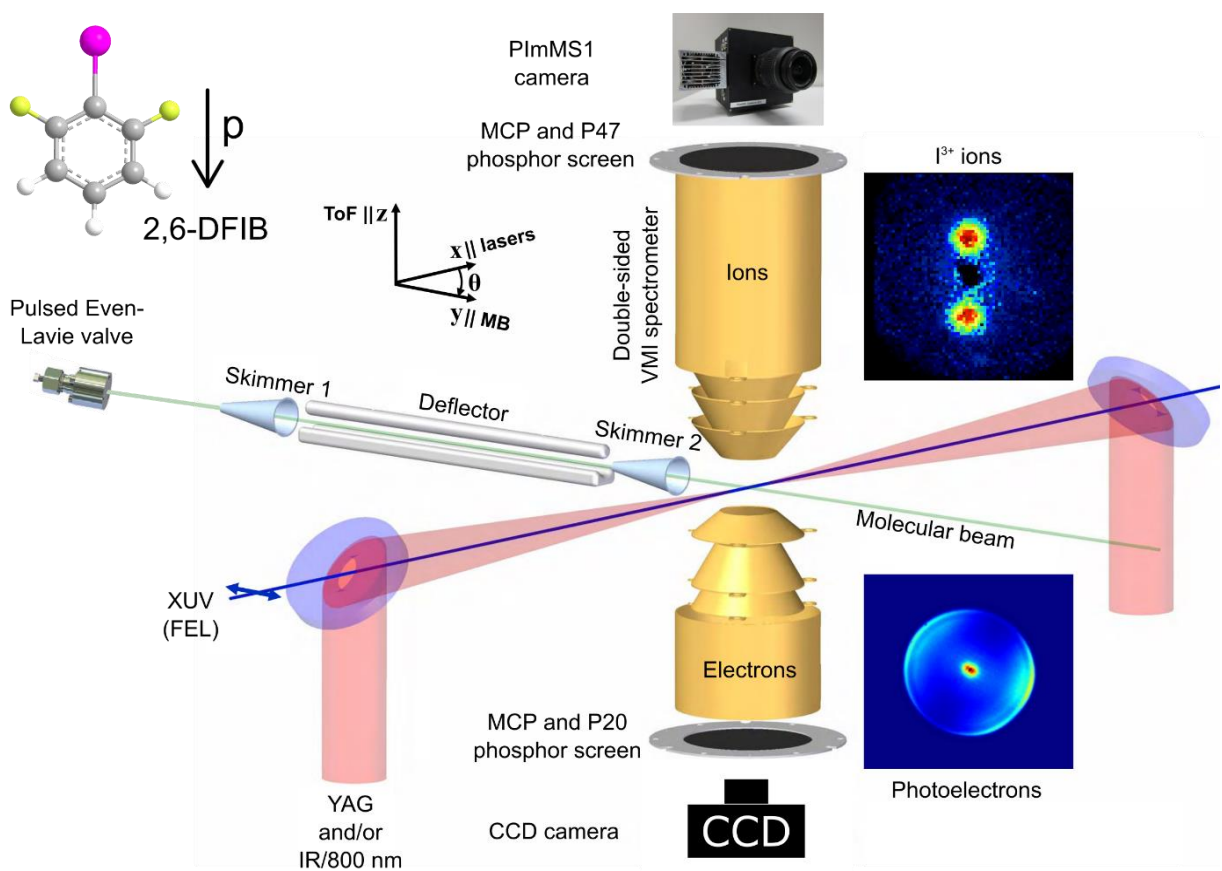

**Figure 1:** Experimental set-up showing the pulsed molecular beam with the electrostatic deflector, the laser beam paths of the alignment and probe lasers, and the double-sided velocity map imaging spectrometer used to simultaneously image the recoil velocities of both the ions and electrons from the molecule. The ions were detected with an MCP and a fast P47 phosphor screen detector coupled with a PImMS1 camera. The electrons were detected with an MCP and a P20 phosphor screen detector coupled with a commercial CCD camera but are not discussed further in this paper. Typical ion and electron images are shown on the right. A model of the 2,6-DFIB molecule is shown on the top left, with the direction of the molecule's electric dipole moment depicted by a black arrow.

The kinetic energies and momentum distribution of the resulting (positive) fragment ions were imaged using a double-sided velocity map imaging spectrometer [*Strüder 2010, Rolles 2014*] equipped with 75-mm microchannel plates and a fast P47 phosphor screen detector, as shown in Fig. 1. The light emitted from the phosphor screen was recorded with the Pixel Imaging Mass Spectrometry (PImMS) camera [*Nomerotski 2010, John 2012, Sedgwick 2012, Brouard 2012, Amini 2015*], a time-stamping camera that records the hit position and arrival time of up to four events per pixel in a given time-of-flight cycle. For each pixel, events were stored in one of four registers, which were read out on a shot-by-shot basis. This single-shot data is saved along with a unique FEL pulse ID that allows correlating the PImMS data with various XUV pulse parameters determined by the FLASH diagnostics system [*Savelyev 2017*]. The experiments reported here were performed with the PImMS1 version of the camera, which records each event with a 12.5 ns timing precision, and comprises an array of 72 x 72 pixels. The PImMS camera was also operated at 10 Hz, and the internal clock of the sensor was electronically synchronized to the master clock of the FEL. In parallel to the ion detection with the PImMS camera, the electrons produced by the interaction of the FEL and the laser pulses with the DFIB molecules were detected at the other side of the double-sided VMI spectrometer using an MCP and a P20 phosphor screen detector coupled with a commercial CCD camera. A typical photoelectron spectrum produced by the ionization of DFIB with FEL pulses is shown in Fig. 1, but no further analysis or interpretation of the electron data is attempted in this paper.

## 3    Results and Discussion

### 3.1    One-Dimensional Alignment

Ion images of $F^+$, $I^+$, $I^{2+}$, and $I^{3+}$ following the ionization of DFIB molecules using either 800 nm NIR photons or 11.6 nm XUV photons are shown in Figure 2, with and without the presence of the adiabatic alignment field. Without the alignment pulse, the momentum distributions of the iodine ions created by ionization of the molecule with the FEL pulse are isotropic, while they exhibit two highly localized maxima along the polarization direction of the Nd:YAG pulse when the molecules are aligned. For the DFIB molecule, the most polarizable axis is parallel to the C-I axis, which will therefore align along the polarization direction of the Nd:YAG pulse.

The degree of molecular alignment is characterized experimentally by calculating the expectation value $<\cos^2 \theta_{2D}>$, where $\theta_{2D}$ is the angle between the projection of the recoil velocity of the detected iodine ion on the detector plane and the polarization axis. With our experimental conditions, we observed a degree of alignment of up to $<\cos^2 \theta_{2D}>$ = 0.97 for horizontal polarization of the NIR probe pulses, 0.93 for vertically polarized NIR probe pulses (not shown), and 0.94 for horizontally polarized XUV probe pulses. These values are in good agreement with those reported previously [*Nevo 2009*]. We do not observe a significant systematic variation of $<\cos^2 \theta_{2D}>$ for the different charge states of the iodine ions. The fact that the alignment can be probed with high fidelity using both NIR and XUV pulses suggests that for the C-I cleavage, the "axial recoil approximation" [*Zare 1967, Zare 1972*] is valid in both cases, i.e. the emission direction of the iodine ions reflects the direction of the C-I axis at the moment the initial ionization occurs. However, we would like to point out that when determining the degree of alignment via strong-field ionization with a femtosecond NIR laser, the polarization direction of the probe laser pulses is usually set perpendicular to the detector plane. This is done in order to avoid so-called "geometric alignment" (or probe) effects due to the enhanced ionization and fragmentation probability for molecules with the C-I axis parallel to the polarization direction of the probe pulse, which lead to higher values of $<\cos^2 \theta_{2D}>$ and thus overestimate the degree of alignment that was achieved. This precaution is typically not necessary when probing via (non-resonant) inner-shell ionization [*Küpper 2014*], since the ionization probability is almost isotropic, as can be seen from the 2D momentum distributions shown in the top row of Fig. 2. These all yield $<\cos^2 \theta_{2D}>$ = 0.5, despite the fact that polarization of the FEL pulse is parallel to the detector, i.e., vertical in the images in Fig. 2. Thus, the FEL pulse provides an unbiased probe of the spatial alignment of molecules independent of its polarization direction.

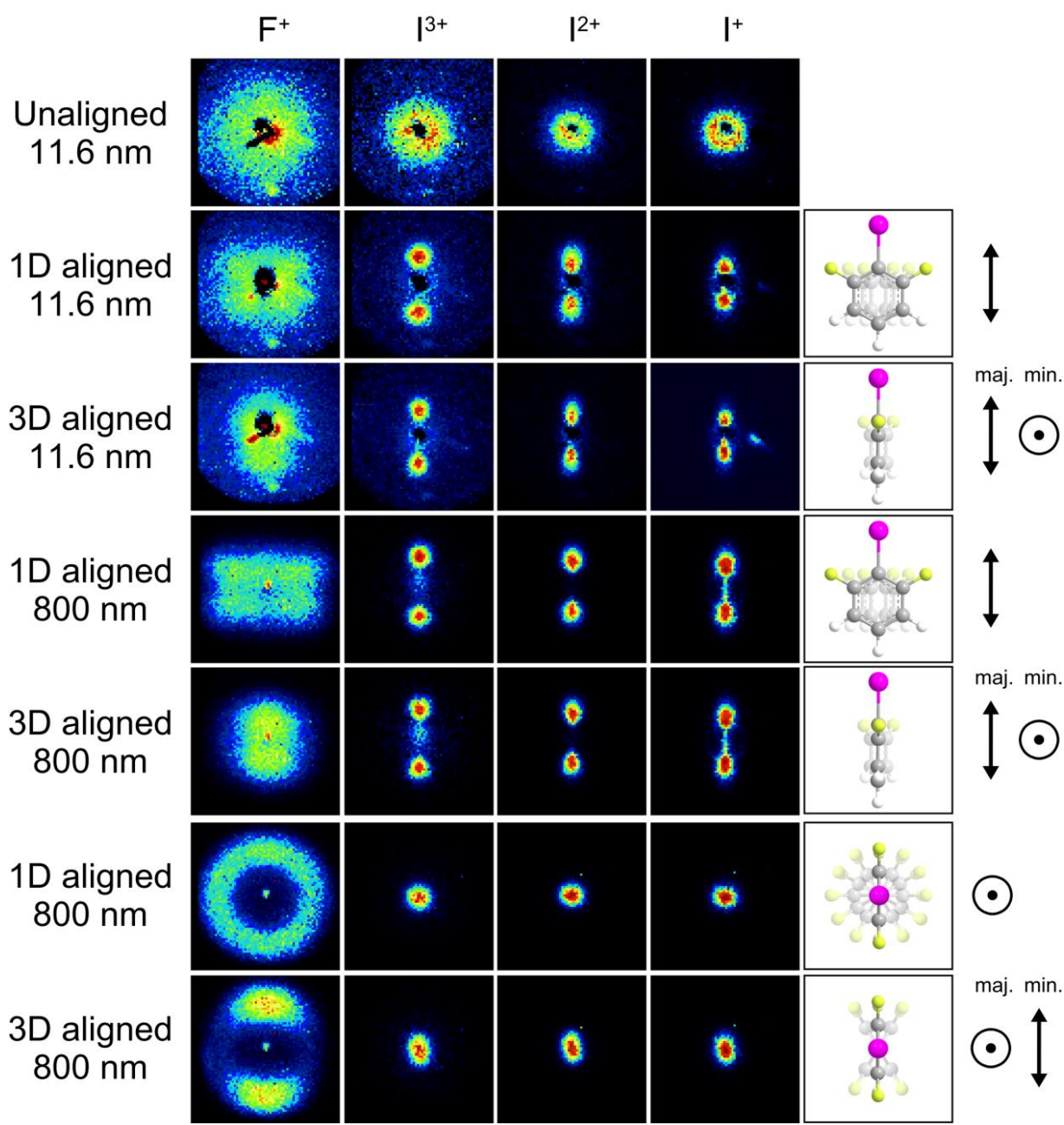

**Figure 2:** Velocity map images of $F^+$, $I^+$, $I^{2+}$, and $I^{3+}$ ions following the ionization of 2,6-difluoroiodobenzene (DFIB) using either 11.6 nm or 800 nm photons recorded with the PImMS camera. Background images recorded with the molecular beam switched off were subtracted. The first row of images was recorded without the adiabatic alignment field. For the subsequent rows, the direction of the (major and minor) polarization axes of the Nd:YAG pulse and the resulting molecular orientation are sketched on the right. Note that the images with 11.6 nm and 800 nm were taken with different spectrometer voltages (see text). The dark spot close to the centre of the images taken with the XUV pulse is due to saturation of the corresponding PImMS camera pixels.

Next, we discuss the $F^+$ ion images, which can be used to characterize the alignment of the molecular plane in space [*Viftrup 2007, Nevo 2009, Ren 2014*]. In the case of 1D alignment by linearly polarized Nd:YAG laser pulses, free rotation about the C-I axis causes the $F^+$ ions to be emitted into two torus-shaped distributions around the alignment axis, which are projected onto the two-dimensional detector either as two thick lines or as a ring, depending on the direction of the polarization axis with respect to the detector. These features are particularly evident in the images taken with the NIR probe pulse, while the images taken with the XUV probe pulse show the same features but slightly less pronounced. This may be due, in parts, to more contamination from isotropic background ions in the XUV case, which we were not able to subtract completely. We also note that the images for the XUV probe pulse were taken at about a factor of two higher extraction voltages on the VMI spectrometer, which reduces the flight time and hence the radius of the measured images, thus causing the momentum distributions to be more compressed than for the NIR case. The dark spot that can be seen close to the centre of the images taken with the XUV pulse is due to saturation of the corresponding PImMS pixels by $Ne^+$ (from the carrier gas) and $H_2O^+$ (from residual gas) ions. These two peaks contain significantly more counts in the case of ionization with XUV pulses than for ionization with NIR pulses, since their ionization cross-section is much higher in the XUV regime than at the NIR intensity used here.

The capability of PImMS to record ion images for several fragments simultaneously also allows for the analysis of angular covariances, as demonstrated in a previous study [*Slater 2014*]. Unfortunately, we are unable to observe such covariances in the present data since we had lowered the voltage across the MCP detectors in order to avoid saturation of the camera and thus inadvertently reduced the detection efficiency to a point where the probability of detecting two ions emitted from the same molecule was negligible. Nevertheless, the possibility of covariance analysis is a major advantage of using the PImMS camera for VMI experiments if camera saturation can be avoided by lower intensities and/or a more dilute target rather than by lowering the detector voltages and images. In that case, approximately 20,000 laser shots are required to observe angular covariances with statistical significance. Note that for a fluctuating light source, a partial-covariance analysis is necessary, which requires a higher number of laser shots.

## 3.2   Three-Dimensional Alignment and Orientation

Three-dimensional alignment of the DFIB molecules was achieved with an elliptically polarized Nd:YAG laser pulse with a corresponding ellipticity intensity-ratio of 2.2 : 1. For an elliptically polarized laser pulse, the most polarizable axis of the molecule, i.e. the C−I axis, aligns along the major axis of the elliptic field, whereas the molecular plane is confined in the polarization plane [*Viftrup 2007, Nevo 2009, Ren 2014*]. The latter can be observed in the $F^+$ ion images shown in Fig. 2. This is particularly pronounced if the major axis of the elliptically polarized field is oriented perpendicular to the ion detector plane, as shown in the bottom two rows of Fig. 2. For that configuration, the ion images present an "end view" of the molecule, where the $I^{n+}$ ions are strongly confined to a central spot, while the $F^+$ ions, in the case of 1D alignment, are projected onto a ring, since the molecular plane is not fixed and can rotate freely around the C−I axis. If, instead, an elliptically polarized pulse is used for alignment, the $F^+$ ions show a pronounced localization along the minor axis of polarization, indicating that the molecular plane is now confined to the plane defined by the elliptical polarization.

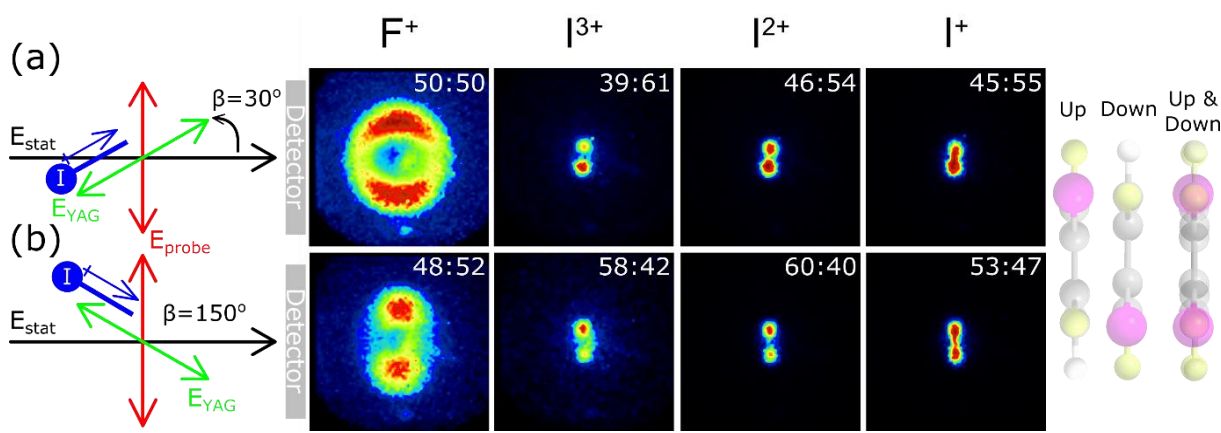

**Figure 3:** $F^+$, $I^{3+}$, $I^{2+}$, and $I^+$ ion images after the Coulomb explosion of 2,6-difluoroiodobenzene molecules using 800-nm photons. In (a), the molecules were 1D-aligned and oriented with the I-end downwards using a linearly polarized light that was rotated away from the detector plane by an angle of β = 30°. In (b), the molecules were 3D-aligned and oriented with the I-end upwards using elliptically polarized light that was rotated away from the detector plane by an angle of β = 150°. The sketches on the left hand side show the polarization axis of the probe (red), of the YAG alignment pulse (green), and the direction of the static electric field (black). The orientation of the molecule and its permanent dipole moment are shown in blue. A sketch of the corresponding molecular orientation, as seen from the perspective of the ion detector, is shown on the right. The calculated up-down asymmetry is indicated in the top right corner of each image.

If the polarization direction of the Nd:YAG laser pulse is parallel to the detector plane, the molecules are aligned perpendicular to the direction of the static electric field inside the VMI spectrometer, and no preferential orientation of the molecule is introduced. The ion images, therefore, show equal intensities in the upper and lower halves of the detector image. However, if the polarization direction (or major axis of polarization) is rotated such that it has an angle other than 90° with respect to the static field (see sketch in Fig. 3), the interaction between the permanent dipole moment of the DFIB molecule within the static electric field of the spectrometer combined with the laser field will induce an orientation of the molecules [*Friedrich 1999*]. The permanent dipole moment in DFIB (p = 2.25 D) points from the iodine end (negative end) to the phenyl ring (positive end) [*Nevo 2009*], as sketched in Fig. 1. Therefore, the iodine end of the molecule orients preferentially towards the positive direction of the electric field, which is away from the ion detector. In the corresponding iodine ion images shown in Fig. 3, taken for angles of 30° and 150° between the major axis of polarization and the electric field of the VMI, this leads to an up-down asymmetry that can be used to quantify the degree of orientation achieved. The corresponding asymmetry ratios, $I_{up}/(I_{up} + I_{down})$ : $I_{down}/(I_{up} + I_{down})$, obtained by integrating the counts in the upper and lower halves of the detector image, are indicated in the top right corner of the images. The data show that we achieved an orientation ratio of around 60:40 in this geometry, with the values obtained from the $I^+$ images being slightly lower due to saturation of the central pixels in those images. In principle, the orientation effect should also be visible in the time-of-flight spectrum, with the $I^{n+}$ peaks splitting into a "forward" and "backward" peak corresponding to the ions emitted towards or away from the ion detector [*Nevo 2009*]. However, the time-resolution in the present experiment was not sufficient to resolve this splitting.

Note that in this particular geometry, the $F^+$ ion distributions does not show an asymmetry due to the orientation of the molecules. This can be understood if we recall that the ion images for vertical polarization of the Nd:YAG pulse (or for an angle close to vertical polarization) corresponds to the end-view of the molecule, as sketched on the right-hand side of Fig. 3. In that case, the two maxima in the $F^+$ ion image correspond to the fluorine atoms on either side of the molecule (i.e. bound to positions 2 and 6 in the molecule) and *not* to the fluorine atoms pointing towards or away from the ion detector.

### 3.3 Coulomb Explosion after Strong-Field and Inner-shell Ionization

Figure 4 shows the ion time-of-flight mass spectra of 1D-aligned DFIB molecules after strong-field ionization with an intense 800-nm pulse (black line) and with an intense 11.6-nm XUV pulse (green line), extracted from the same PImMS data set as the velocity-map images shown in Fig. 2. The figure also includes the mass spectra for unaligned DFIB molecules (red line) compared to the direct readout of the decoupled analog MCP signal with a high-resolution high-speed digitizer (*Acqiris U1065A*, blue line) following ionization by the XUV pulse. Although the mass spectrum extracted from the MCP signal has significantly better resolution, the mass spectrum obtained with the PImMS camera resolves most of the relevant fragments, especially for *m/q* > 30. At lower *m/q*, the resolution of the PImMS spectrum is not sufficient to resolve individual mass peaks. The time-of-flight resolution may be improved by centroiding the PImMS data [*Slater 2014*] in position and time. We did not do this for the present data, since accurate centroiding requires that each ion yields a signal with a spatial extent of several pixels on the PImMS sensor, and this was not the case at the low MCP detector gains used. A further limitation of the time-of-flight spectra recorded with the PImMS camera is that fragments ejected with very low kinetic energies, such as $OH^+$, $H_2O^+$, and $Ne^+$, are suppressed due to the saturation of the central few pixels.

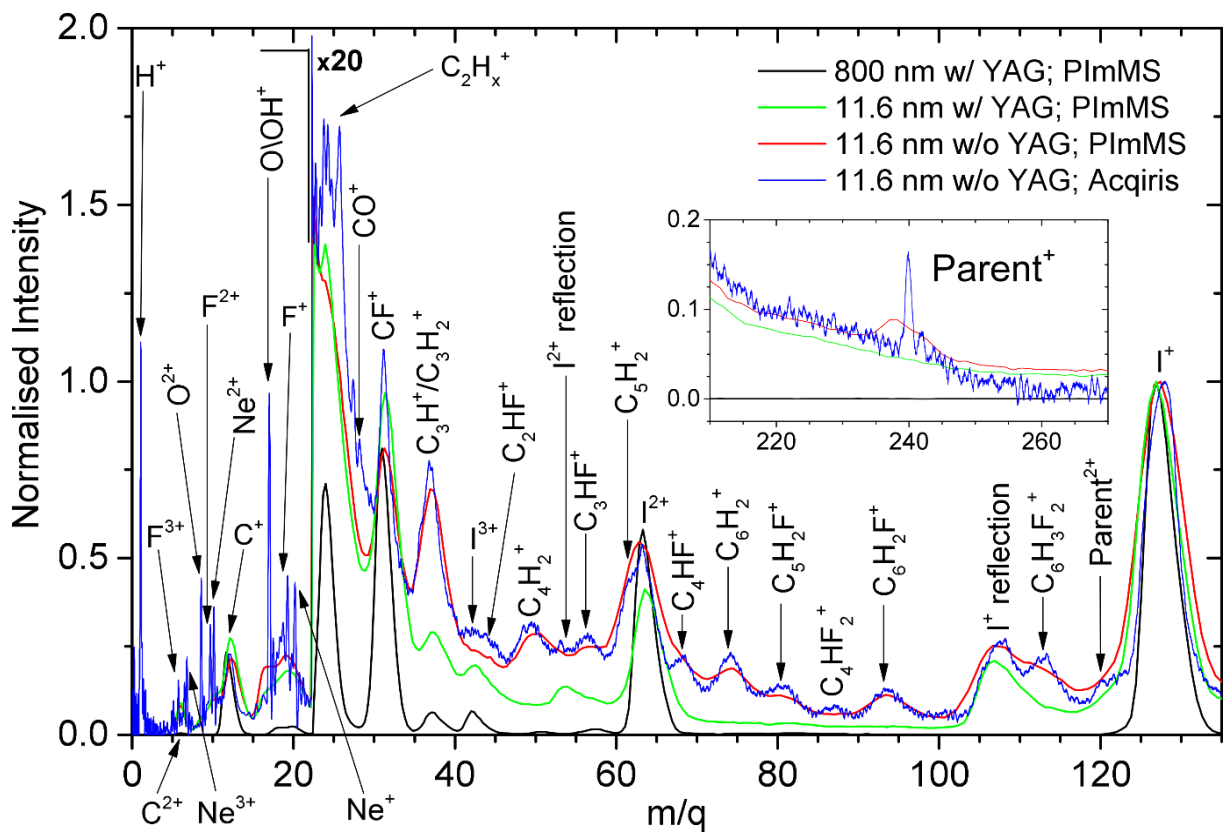

**Figure 4:** Ion time-of-flight mass spectra of DFIB recorded with the PImMS camera after strong-field ionization with an intense 800-nm pulse (black line) and with an intense 11.6-nm XUV pulse (green and red lines). The XUV spectra are shown with (green) and without (red) the presence of the Nd:YAG alignment laser pulse, and are also compared to the spectrum obtained from directly reading out of the decoupled analog MCP signal with a high-resolution digitizer (blue). The spectra are normalized to the maximum of the $I^+$ peak and have been scaled up by a factor of 20 for $m/q > 22$. The inset shows the $m/q$-region of the parent ion.

Comparing the spectra recorded with and without the Nd:YAG pulse (red and green line, respectively), we can clearly observe that the Nd:YAG pulse destroys most of the larger molecular fragments including the parent ion (see inset in Fig. 4), as we have already observed previously for another fluorobenzene derivative [*Boll2014*], resulting in a spectrum that is dominated by $I^{n+}$ (with $n=1,2,3$) and small carbon-containing fragments. Furthermore, despite the fact that the ionization pathways for strong-field ionization and (multi-photon) inner-shell ionization are quite different, with the former sequentially removing valence electrons from the molecule, while the latter proceeds via a sequence of inner-shell ionizations followed by Auger decay, the corresponding ion mass spectra (black and green lines) are quite similar at the given intensities in this experiment. The large contribution of $I^{2+}$ and $I^{3+}$ in the 11.6-nm spectrum is a clear indication of multi-photon processes in the XUV ionization, since fragmentation spectra taken with synchrotron radiation at the same photon energy show only a small signal from $I^{2+}$ ions and no $I^{3+}$ ions [*Ablikim 2017*]. The spectra taken following 11.6-nm excitation also show a larger contribution from $F^+$, $F^{2+}$, and small carbon containing fragments, suggesting a generally higher degree of ionization than with the 800-nm pulses at the given intensity. Note that in the 11.6-nm spectra, which were recorded at higher extraction potentials, the $I^+$ and $I^{2+}$ peaks are each accompanied by a smaller artefact peak at slightly smaller *m/q* (corresponding to shorter flight times), labelled "reflection". These peaks appear when the drift tube on the ion side of the spectrometer is at a more negative potential than the front of the MCP. In this case, secondary electrons created by ion impact on the mesh that terminates the drift tube are accelerated towards the MCP, where they are detected at slightly shorter flight times than the corresponding ions.

Further information on the Coulomb explosion process can be obtained by analysing the fragment ion kinetic energies. For this purpose, the ion images of 1D-aligned DFIB shown in Fig. 2 are inverted using the pBasex package [*Garcia 2004*], and the resulting radial distributions are converted to a kinetic energy scale using a simulation of the VMI spectrometer performed in SIMION [*SIMION81*]. The kinetic energy calibration was cross-checked by comparing measurements of the kinetic energies resulting from the UV-induced dissociation of $CH_3I$ with known values from the literature as well as by comparing the results for the XUV-induced Coulomb explosion of DFIB with similar data obtained with synchrotron radiation [*Ablikim 2017*]. On the basis of these cross-checks, the uncertainty of the energy calibration, $\Delta E/E$, was determined to be less than 5% for ion kinetic energies above 1 eV and less than 10% for energies below 1 eV. The resulting kinetic energy distributions for iodine fragments of 1D-aligned DFIB are shown in Figure 5. A clear increase in the ion kinetic energy release is seen upon increasing ion charge. Furthermore, for the given FEL and NIR-laser intensities and pulse lengths, the maxima of the $I^+$ and $I^{2+}$ kinetic energy distributions are rather similar for strong-field ionization with 800 nm and for inner-shell multi-photon ionization with 11.6 nm. This suggests that the kinetic energies are rather independent of the details of the ionization process and mostly depend on the total charge state of the molecule upon breakup. In this context, it is interesting that the $I^{2+}$ kinetic energy distribution obtained following 11.6-nm ionization extends to higher energies than the corresponding distribution for 800-nm ionization, while the $I^{3+}$ kinetic energy distribution is shifted to higher energies altogether. This confirms the above finding of a higher overall degree of ionization in the XUV case. In order to quantify this further, we can compare the experimental kinetic energies with the energies expected from a classical Coulomb explosion model. The vertical lines at the top of the plot indicate the calculated kinetic energies of $I^+$ (black vertical lines), $I^{2+}$ (red vertical lines), and $I^{3+}$ (blue vertical lines) fragments assuming purely Coulombic repulsion between the iodine ion and a single co-fragment of varying charge. For the distance between the charges on the iodine fragment and the co-fragment, a value of 487.3 pm (corresponding to the distance between the iodine atom and the furthest-away carbon atom in the equilibrium geometry of the molecule) was chosen, which gives the best agreement with the experimentally determined kinetic energy release of the $I^+$ - $C_6H_3F_2^+$ channel, obtained from an electron-ion-ion coincidence experiment performed with synchrotron radiation at the same photon energy [*Ablikim 2017*].

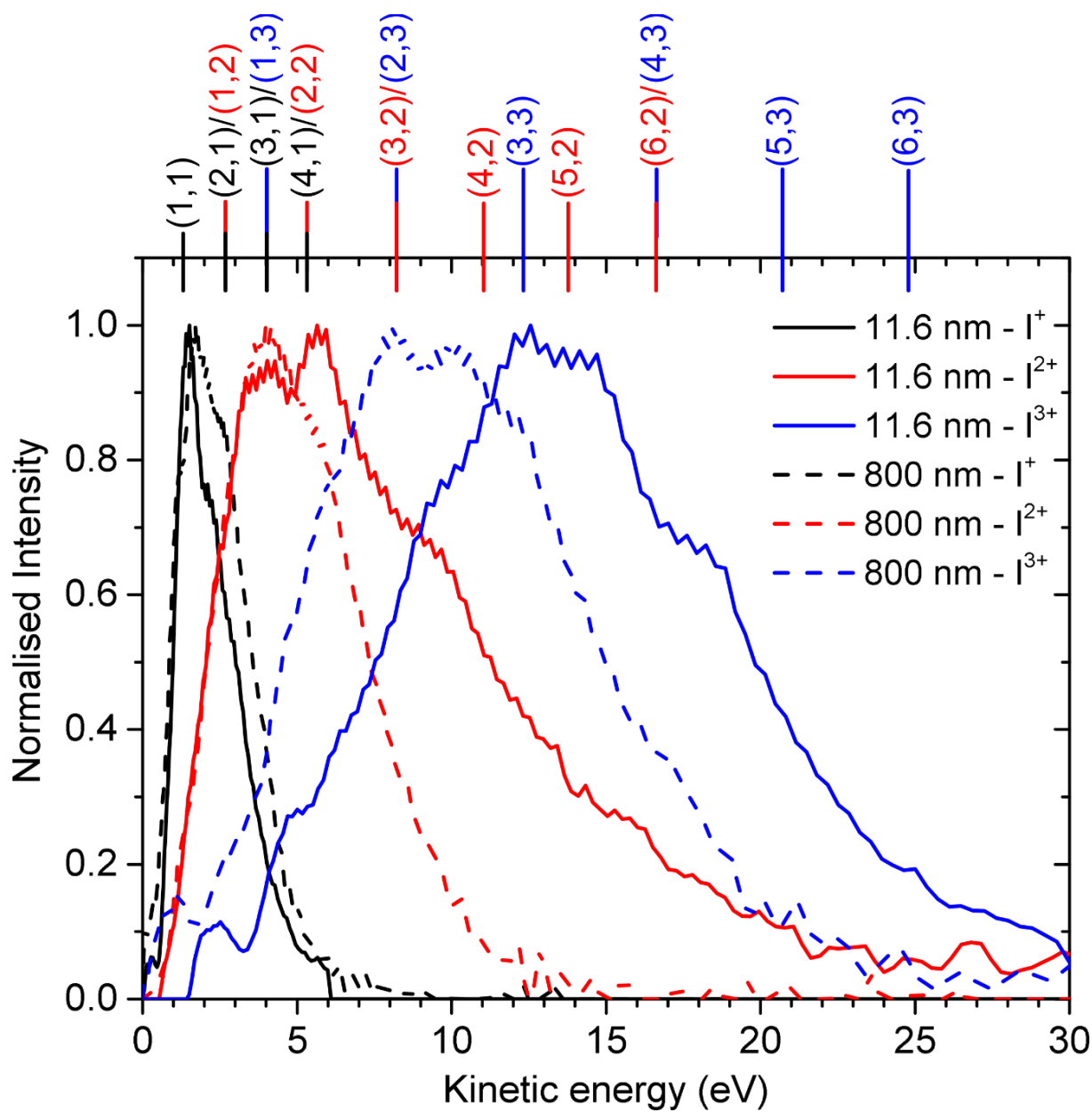

**Figure 5:** Kinetic energy distributions of $I^+$ (black), $I^{2+}$ (red), and $I^{3+}$ (blue) ions from 1D-aligned DFIB molecules extracted from the corresponding pBasex-inverted ion images for multi-photon inner-shell ionization using an 11.6-nm XUV pulse (solid lines) and strong-field ionization using an 800-nm NIR pulse (dashed lines). Each distribution was normalized to its maximum value. The vertical lines at the top of the figure indicate the expected kinetic energy of the $I^+$ (black), $I^{2+}$ (red), and $I^{3+}$ (blue) ions assuming pure Coulombic repulsion from a singly, doubly, triply, etc. charged co-fragment (see text). They are labeled $(n,m)$, where $m$ is the charge of the iodine ion and $n$ the charge of the co-fragment.

Using these model calculations, we can assign the main peak and the high energy shoulder in the $I^+$ kinetic energy distributions for both 11.6-nm and 800-nm pulses as being due to fragmentation with a singly and doubly charged co-fragment, respectively. This confirms the earlier tentative assignment of the two features that were previously observed in the $I^+$ images obtained after strong-field ionization [*Viftrup 2007, Nevo 2009, Ren 2014*]. Similarly, the $I^{2+}$ kinetic energy distribution obtained after strong-field ionization is dominated by contributions stemming from singly and doubly charged co-fragments, while it clearly contains contributions from triply, quadruply, and maybe even more highly charged partners in the XUV case. For $I^{3+}$, the kinetic energy distribution indicates fragmentation with mainly doubly and triply charged co-fragments in the 800-nm case, and with triply and quadruply co-fragments in the XUV case. Note that while our model calculations do not differentiate further how the charges are distributed on the co-fragment(s), our mass spectra clearly show that the molecule eventually breaks into several smaller moieties, as seen in Fig. 4.

## 4    Conclusions

We have reported the first application of the PImMS camera for velocity-map ion imaging at a free-electron laser. The PImMS sensor is able to record simultaneously the time-of-flight spectrum and the 2D momentum distributions of all ionic fragments and can cope with count rates of several hundred ions per shot, thanks to the ability to detect up to 4 charged particles per pixel and per detection cycle. The ability to detect the ion images for all fragments simultaneously is of particular advantage for time-dependent pump-probe studies that we will present in forthcoming publications. Combining the PImMS camera with velocity-map ion imaging and with laser-induced molecular alignment thus represents an important step towards the investigation of ultrafast molecular dynamics using XUV-induced Coulomb explosion imaging experiments.

In the present work, we have demonstrated a high degree of one- and three-dimensional alignment and orientation, obtained by laser-induced adiabatic alignment and mixed-field orientation of quantum-state-selected DFIB molecules, and probed using velocity-map ion imaging coupled with the PImMS camera. We have also shown that molecular alignment and orientation can be probed with high fidelity using both strong-field ionization and multi-photon inner-shell ionization and that both lead to similar Coulomb explosion processes despite the very different ways in which the ionization occurs. Non-resonant inner-shell ionization has the advantage that it is typically almost isotropic with respect to the angle between the polarization direction and any molecular axis, which mostly avoids any "geometric alignment" effects that have to be considered when using strong-field ionization as a probe. However, since the fragmentation typically occurs after an intermediate Auger decay process, it may be even more critical to check that the fragmentation indeed occurs as strictly along the direction of the bond axes as in the case of strong-field ionization. In the present case, this seems to be true for the iodine fragments, while it is not as clear for the fluorine fragments, as demonstrated also by recent ion-ion coincidence experiments [*Ablikim 2017*].

**Acknowledgements**

We gratefully acknowledge the work of the scientific and technical team at FLASH, who has made these experiments possible. The support of the UK EPSRC (to M.Br. and C.V. via Programme Grants EP/G00224X/1 and EP/L005913/1), the EU (to M.Br. via FP7 EU People ITN project 238671 and to J.K., P.J., H.S, and D.R. via the MEDEA project within the Horizon 2020 research and innovation programme under the Marie Skłodowska-Curie grant agreement No 641789), STFC through PNPAS award and a mini-IPS grant (ST/J002895/1), and a proof of concept grant from ISIS Innovation Ltd. are gratefully acknowledged. We also acknowledge the Max Planck Society for funding the development of the CAMP instrument within the ASG at CFEL. In addition, the installation of CAMP at FLASH was partially funded by BMBF grant 05K10KT2. K.A. thanks the EPSRC, Merton College, Oxford University and RSC for support. A.L. thanks the DFG via Grant No. La 3209/1-1 for support. N.B., A.Ru., and D.R. acknowledge support from the Chemical Sciences, Geosciences, and Biosciences Division, Office of Basic Energy Sciences, Office of Science, U.S. Department of Energy, Grant No. DE-FG02-86ER13491 (Kansas group) and DE-SC0012376 (U Conn group). D.R., E.S., R.B., C.B, and B.E. were also supported by the Helmholtz Gemeinschaft through the Helmholtz Young Investigator Program. J.K. was, in addition to DESY, supported by Helmholtz Networking and Initiative Funds, by the excellence cluster ``The Hamburg Center for Ultrafast Imaging—Structure, Dynamics and Control of Matter at the Atomic Scale'' of the Deutsche Forschungsgemeinschaft (CUI, DFG-EXC1074), and, with T.K., by the Helmholtz Virtual Institute 419 "Dynamic Pathways in Multidimensional Landscapes". S.Te. is grateful for support through the Deutsche Forschungsgemeinschaft, project B03/SFB755 and project C02/SFB1073. P.J. acknowledges support from the Swedish Research Council and the Swedish Foundation for Strategic Research. A.Ro. is grateful for support through the Deutsche Forschungsgemeinschaft project RO 4577/1-1.